\documentclass[twocolumn,showpacs,preprintnumbers,amsmath,amssymb,floatfix]{revtex4}
\usepackage{graphicx}% Include figure files
\usepackage{dcolumn}% Align table columns on decimal point
\usepackage{bm}% bold math
\usepackage{color}
\def\la{\langle}
\def\ra{\rangle}
\def\beq{\begin{equation}}
\def\eeq{\end{equation}}
\def\be{\begin{eqnarray}}
\def\ee{\end{eqnarray}}

\begin{document}
\title{
Advantage of U+U over Au+Au collisions at constant beam energy 
}
\author{C. Nepali}
\author{G. Fai}
\author{D. Keane}
\affiliation{Center for Nuclear Research, Department of Physics \\
Kent State University, Kent, OH 44242, USA}
\date{\today}
\begin{abstract}
Collisions of deformed uranium nuclei are studied in a Monte-Carlo Glauber model.   
For U+U at zero impact parameter ($b=0$) in the most favorable orientation 
(tip-to-tip), the transverse particle density (charged-particle rapidity density 
per weighted transverse area of the initial 
participant zone) increases by about 35\% compared to Au+Au at $b=0$.  
To estimate the advantage of U+U over Au+Au in the context of real experiments 
at the Relativistic Heavy Ion Collider, we examine the effect of a range of 
centrality cuts on the event sample.  In terms of the transverse particle density, 
the predicted advantage of U+U is about 16\%. 
\end{abstract}
\maketitle
\vspace{1cm}
%
%%%%%%%%%%%%%%%%%%%%%%%%%%%%%%%%%
% 
\section{Introduction}

Large space-time volumes with the highest energy density to 
date, over 15 GeV/fm$^3$ in typical estimates\cite{Gyulassy:2004zy},
have been produced in $\sqrt{s_{NN}} = 200$ GeV Au+Au collisions 
at the Relativistic Heavy Ion Collider (RHIC) 
\cite{Arsene:2004fa,Adcox:2004mh,Back:2004je,Adams:2005dq}.
At such high energy densities, strongly interacting matter is predicted 
%by lattice QCD
to be in a phase commonly referred to as quark-gluon plasma (QGP) 
\cite{Karsch:2004ti}.  A goal of the RHIC program is to study this 
phase in the extended systems created in relativistic collisions of 
heavy nuclei.  

The presence of QGP manifests itself in the equation of state, which 
(together with the initial conditions) determines various measured 
properties and serves as an input to fluid-dynamical calculations.  
It is reasonable to expect that, barring calculational problems, 
the main features of the equation of state can be probed by comparing 
fluid-dynamically calculated quantities to data.  It has been argued 
recently that the quark-gluon matter produced in 200~GeV 
central Au+Au collisions behaves as a perfect fluid with negligible 
viscosity\cite{Hirano:2005wx,Tannenbaum:2006ch}.  
This conclusion was based in part on the 
agreement \cite{Adams:2005dq} of the flow quantity $v_2/\epsilon$ (see 
Eqs.~(\ref{Fourier}) and (\ref{epsilon}) for a definition) in central 
Au+Au with the prediction of an ideal-fluid calculation 
\cite{Kolb:2000sd}. The data seem to indicate that the systems 
produced in 200~GeV central Au+Au collisions have just reached a 
large enough energy density in a sufficient space-time volume for 
the approximate agreement with the fluid-dynamical results.  If 
energy density could be increased even further, a crucial test would 
be whether $v_2/\epsilon$ continues to grow, or whether it saturates 
at the value for an ideal fluid.  

A promising avenue to increase the energy density and/or increase the 
volume of high-density matter without any increase in beam energy is to 
collide heavy deformed (prolate) nuclei.  This has been proposed by 
several theorists \cite{Shuryak:1999by,Li:1999be,Heinz:2005xd}.  Central 
collisions with the long axes aligned with the beam (we call this 
the tip-to-tip configuration, as in Ref.~\cite{Li:1999be}) 
would be the most desired configuration.  However, as long as no 
beams of aligned deformed nuclei are available \cite{Fick:1981hq}, the 
desired configurations have to be selected by experimenters using some 
combination of triggering and offline event selection.  
Recently, Heinz and Kuhlman advocated the use of full-overlap 
collisions between deformed uranium nuclei \cite{Heinz:2005xd}.  
The selection of full-overlap collisions using Zero-Degree Calorimeters 
(ZDCs) was examined in Ref.~\cite{Kuhlman:2005ts}.  
However, various fluctuations and background signals in the 
detectors mean that an event sample with the lowest ZDC signal will 
still include a substantial fraction of collisions in which full 
overlap did not occur.  In this work, we address these issues with 
the help of data on the relevant performance of the STAR detector 
\cite{STAR-NIM}. 

The paper is organized as follows: in Sec.~\ref{calc} we briefly review 
the relevant quantities and outline our calculation, while in Sec.~\ref{res}, 
we present results without and with detector smearing.  In Sec.~\ref{concl}, 
we briefly summarize our findings.
%
%%%%%%%%%%%%%%%%%%%%%%%%%%%%%%%%%
% 
\section{Background and Calculational framework}
\label{calc} 

The elliptic flow anisotropy $v_2$, defined as the second Fourier coefficient 
in the expansion
\beq
\frac{dN}{d\Phi} = A \left[ 1 + \Sigma \;\; 2 v_n \cos{(n\Phi}) \right]
\label{Fourier}
\eeq
of the azimuthal distribution $dN/d\Phi$ of final state particles is a 
sensitive measure of the success of fluid-dynamical models. Based on 
suggestions in Refs.~\cite{Heiselberg:1998es,Sorge:1998mk,Voloshin:1999gs}
it has become customary to consider $v_2/\epsilon$, the elliptic flow 
scaled by the initial spatial eccentricity \cite{Voloshin:1999gs} 
\beq
\epsilon = \frac{\la y^2 \ra - \la x^2 \ra}{\la y^2 \ra + \la x^2 \ra} \,\,\, .
\label{epsilon}
\eeq
The normalization by $\epsilon$ emphasizes that the final momentum anisotropy 
is driven by the initial geometry.  One advantage of using $v_2/\epsilon$ 
is that the impact-parameter dependence is largely removed from its 
fluid-dynamically calculated value.  This quantity is frequently plotted 
against the transverse particle density $(1/S)\, dN_{ch}/dy$, where $dN_{ch}/dy$ 
is the multiplicity of charged particles per unit rapidity and
\beq
S = \pi~\sqrt{\la x^2 \ra ~ \la y^2 \ra}
\label{area}
\eeq
is the transverse area of the overlap zone 
weighted by the number of wounded nucleons \cite{Heiselberg:1998es,Alt:2003ab,Adams:2005dq}. 
The transverse particle density $(1/S)\, dN_{ch}/dy$ can be 
interpreted as a measure of the initial entropy density in the transverse 
plane \cite{Heinz:2005xd}. An increase in this 
quantity means increased particle production, and is thus associated with an 
increase in energy density in the system.  The charged particle multiplicity 
per unit rapidity $dN_{ch}/dy$ can be increased by increasing the number of 
binary collisions, achievable by tip-to-tip collisions of heavy deformed 
nuclei. Furthermore, the U+U tip-to-tip configuration has about the same 
overlap area $S$ as Au+Au collisions at $b=0$, and thus contributes to a 
further increase in $(1/S)\, dN_{ch}/dy$ beyond what would be obtained with 
hypothetical spherical nuclei of the same mass as uranium.  On the other 
hand, for $b=0$ collisions, $S$ increases by $\approx$ 24\% between Au+Au 
and U+U when the latter are in the body-to-body orientation, i.e., when the 
long axes of the nuclei are parallel to each other, but perpendicular to the 
beam.  Thus it can be concluded that even in the experimentally unrealistic 
limit where full-overlap U+U collisions can be isolated, it is necessary to 
also distinguish between different full-overlap configurations in order to 
realize the full potential advantage of U+U collisions. 

The RHIC data as a function of $(1/S)\, dN_{ch}/dy$ reach the perfect-fluid 
value of $v_2/\epsilon \approx 0.2$ in the most central 200 GeV Au+Au 
collisions at around $(1/S)\, dN_{ch}/dy \simeq 28$~fm$^{-2}$.  Since 
viscosity would decrease the predicted $v_2/\epsilon$, %%we can say that 200 GeV 
%% central Au+Au exhausts the fluid-dynamical limiting value of this quantity.   
it would be difficult for a fluid-dynamical model to be reconciled with any 
further increase in the observed ratio $v_2/\epsilon$.  Thus, the behavior 
of this ratio at significantly larger $(1/S)\, dN_{ch}/dy$ should prove 
informative in assessing the applicability of fluid dynamics in central 
collisions at RHIC.  
 
Uranium ($^{238}$U) beams have been proposed as a way to increase 
$(1/S)\, dN_{ch}/dy$ \cite{Shuryak:1999by,Li:1999be,Heinz:2005xd}.  
We assume in this study that, in order to isolate a sample of U+U collisions 
with the maximum possible $(1/S)\, dN_{ch}/dy$, the best available experimental 
procedure is to select events where the fewest possible spectator neutrons are 
detected (low ZDC signal) in coincidence with the highest observed multiplicity 
of produced particles in a broad region centered on mid-rapidity.  Available 
data from the STAR experiment \cite{STAR-NIM} describe this correlation and its 
fluctuation from event to event, and allow the output from the Monte-Carlo 
Glauber model described below to be filtered appropriately to simulate the 
relevant experimental limitations. 

\begin{figure}[h]
\begin{center}
\resizebox{5cm}{!}{\includegraphics*{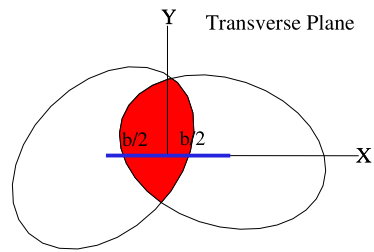}}
\caption{
(Color online) Collision geometry.
\label{fig:geometry}}
\end{center}
\end{figure}

\begin{figure}[h]
\begin{center}
\resizebox{7cm}{5cm}{\includegraphics*{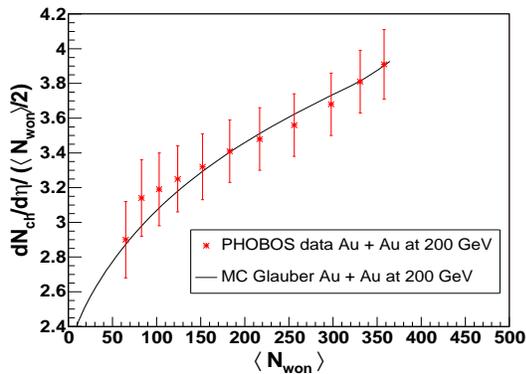}}
\caption{
(Color online) Fit to PHOBOS data for Au+Au at 200 GeV, using Eq.~(\ref{KNpar}). 
\label{fig:KNpar}}
\end{center}
\end{figure}

We represent the quadrupole deformation of the ground-state uranium nucleus in the 
standard \cite{Bohr:1969} way: we take a Saxon-Woods density distribution with 
surface thickness $a = 0.535$ fm and with $R = R_{\rm sp}(0.91 + 0.27\cos ^2\theta)$, 
where $\theta$ is the polar angle relative to the symmetry axis of the nucleus 
and $R_{\rm sp} = 1.12 A^{1/3} - 0.86 A^{-1/3}$ fm \cite{Bohr:1969}. The small 
hexadecapole moment of the uranium nucleus is neglected.  This yields 
$R_{\rm long}/R_{\rm perp} \simeq 1.29$ \cite{Kolb:2000sd,Shuryak:1999by,Heinz:2005xd}.  
The orientation of the first and second nucleus in the colliding pair is fixed 
by the two angles ($\theta_p$, $\phi_p$) and ($\theta_t$, $\phi_t$), respectively. 
The angles $\theta_{p}$ and $\theta_{t}$ describe the orientation of the symmetry 
axis relative to the beam direction, and they are uniformly distributed in 
[0,\, $\pi/2$].  The azimuthal angles $\phi_{p}$ and $\phi_{t}$ describe rotations 
about the beam direction, and they are uniformly distributed in [0,\, 2$\pi$].  A 
schematic view of a cut in the transverse plane illustrates the collision geometry 
in Fig.~\ref{fig:geometry}, where the lines represent equivalent sharp surfaces, 
and the shaded area corresponds to the overlap region in a collision at impact 
parameter $b$ with fixed orientations of both nuclei. 
To simulate a uranium 
nucleus, 238 nucleons are selected randomly according to the appropriate Saxon-Woods 
distributions with distances between any two nucleons satisfying $r_{ij} \geq 0.4$ fm.  
In the collision process, nucleons are considered wounded if the transverse distance 
between them becomes $\leq \sqrt{4.2/\pi}$ fm, where we use 4.2 fm$^{2}$ for the 
nucleon-nucleon cross section at 200 GeV. 
To convert between track densities in rapidity $y$ and in pseudorapidity $\eta$, we 
use the approximation $dN_{ch}/dy \simeq 1.15\, dN_{ch}/d\eta$ \cite{Adler:2002pu}, 
and apply the parameterization
\beq
\frac{dN_{ch}}{d\eta} = n_{pp} [ x N_{b} + (1 -x) N_{w}/2 ] \,\, ,
\label{KNpar}
\eeq
where $N_{b}$ is the numbers of binary collisions and $N_{w}$ is the number of 
wounded nucleons \cite{Kharzeev:2000ph}.  The values of the parameters used are 
$x=0.15$ and $n_{pp} = 2.19$ at 200 GeV.  These values provide a reasonable fit 
to the PHOBOS data \cite{Back:2002uc} as shown in Fig.~\ref{fig:KNpar}.
%%%%%%%%%%%%%%%%%%%%%%%%%%
\section{Results without and with detector smearing}
\label{res}

\subsection{No smearing}
\begin{figure}[h]
\begin{center}
\resizebox{7cm}{5cm}{\includegraphics*{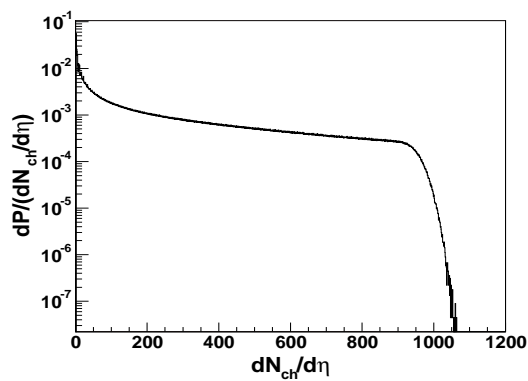}}
\caption{
Charged particle multiplicity distribution for U+U collisions.
\label{fig:mult-no-smear}}
\end{center}
\end{figure}

Figure~\ref{fig:mult-no-smear} displays the charged multiplicity distribution 
$dN_{ch}/d\eta$ for U+U collisions.  Since we average over all possible 
orientations, this distribution is similar to that obtained for hypothetical 
spherical nuclei with mass 238.
For ideal tip-to-tip U+U collisions ($b = 0$ fm, $\theta_{p}$ = 0, $\theta_{t}$ = 0) 
we obtain $(1/S)\, dN_{ch}/dy \simeq$ 42.6 fm$^{-2}$. 
For body-to-body collisions ($ b=0,\, \theta_{p} = \theta_{t} = \pi/2,\, 
\phi_{p} = \phi_{t} = 0$) we have $(1/S)\, dN_{ch}/dy \simeq$ 31.7 fm$^{-2}$,
surprisingly close to 31.5 fm$^{-2}$, the value for Au+Au at $b=0$ fm. 

Although $dN_{ch}/dy$ is larger for U+U than for Au+Au, 
the larger overlap area for body-to-body collisions offsets the increase and 
results in a value of $(1/S)\, dN_{ch}/dy$ close to that for Au+Au.  The 
approximately 35\% increase in $(1/S)\, dN_{ch}/dy$ in tip-to-tip U+U collisions 
compared to Au+Au is due to the increase in $dN_{ch}/dy$, since the overlap areas 
are similar in these cases.
 
Under the best of circumstances, an experiment can hope to select certain ranges 
of angles and impact parameter.  For illustrative purposes, we choose a somewhat 
arbitrary range of definitions for near tip-to-tip configurations: 
$b \leq$ 1 or 2 fm, with $\theta_{p}$ and $\theta_{t}$ both within $10^\circ$ or 
$20^\circ$, and for simplicity we keep $\theta_p = \theta_t$ throughout. 

Even if we impose the strictest definition above for near tip-to-tip events 
($\theta_p = \theta_t \leq 10^\circ$ and $b \leq 1$ fm) and a similarly restrictive 
definition for near body-to-body collisions, almost all of the near tip-to-tip and 
near body-to-body events lie in the top 3\% of the $dN_{ch}/d\eta$ (or, equivalently 
$dN_{ch}/dy$) distribution.  For Au+Au collisions with $dN_{ch}/dy$ in the top 3\% 
of the event sample, the additional requirement that the number of spectator 
nucleons lie in the bottom 1\% of its parent distribution leads to little or no 
further change in $(1/S)\, dN_{ch}/dy$.  For near tip-to-tip 
U+U configurations, there are significant increases of 32--36\% over central 
Au+Au.  Details related to various near tip-to-tip U+U selection 
conditions can be found in Table~I, and as before, a spectator cut has 
negligible additional effect on $(1/S)\, dN_{ch}/dy$.  If a near body-to-body 
configuration is defined as $0 \leq b \leq 1$ fm, $\theta_p,~\theta_t 
\geq 80^\circ$, and $|\phi_p - \phi_t| \leq 10^\circ$, then the mean value of 
$(1/S)\, dN_{ch}/dy$ comes out to be 31.3 fm$^{-2}$, very close to the ideal 
body-to-body case mentioned above, and still negligibly different from the 
value for central Au+Au.  
\begin{table}[h]
\begin{center}
\begin{tabular}{|c|c|c|c|c|c|}
       \hline 
& spectator   &  \multicolumn{2}{c|}{0 $\leq$ b $\leq$ 1 fm}  
              &  \multicolumn{2}{c|}{0 $\leq$ b $\leq$ 2 fm} \\
& count       &  $\%$evt & $\la \frac{1}{S} \frac{dN_{ch}}{dy} \ra$ 
              & $\%$evt & $\la \frac{1}{S} \frac{dN_{ch}}{dy} \ra$ \\
       \hline
$\theta \leq 10^\circ$&    any      &    0.174    &    42.25    &    0.687    &    41.59 \\
       \hline
$\theta \leq 20^\circ$&    any      &    0.707    &    41.75    &    2.756    &    41.13 \\
       \hline
$\theta \leq 10^\circ$& lowest 1$\%$&    0.173    &    42.26    &    0.478    &    41.95 \\
       \hline
$\theta \leq 20^\circ$& lowest 1$\%$&    0.704    &    41.76    &    1.903    &    41.50 \\
       \hline
\end{tabular}
\end{center}
\caption{Mean values of $(1/S)\, dN_{ch}/dy$ and percentage of the event class 
in near tip-to-tip U+U configurations relative to the number of events in the 
top 3\% of $dN_{ch}/dy$.}
\end{table}

\begin{figure}[h]
\begin{center}
\resizebox{7cm}{5cm}{\includegraphics*{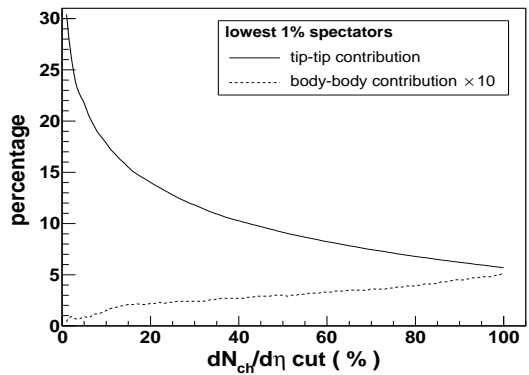}}
\caption{ 
Among events where the spectator nucleon count lies in the lowest 1\% of all 
collisions, the percentage of near tip-to-tip ($\theta \leq 20^\circ, ~b \leq 2$ fm) 
and near body-to-body ($\theta \geq 70^\circ, ~|\phi_p - \phi_t| \leq 20^\circ$, 
~$b \leq 2$ fm) configurations as a function of an additional cut on $dN_{ch}/d\eta$.  
The horizontal scale is normalized such that 100\% corresponds to the full 
sample where only the spectator cut has been applied.
\label{fig:spect-1percent}}
\end{center}
\end{figure}

In the discussion above, central collisions were selected primarily via their high 
multiplicity.  Next, we use a low spectator count as the initial centrality criterion, 
and explore the effect of an additional cut on total multiplicity.  Results are 
presented in Fig.~\ref{fig:spect-1percent}, and it is evident 
that the ratio of near tip-to-tip to near body-to-body events increases very 
substantially as more central collisions are selected.  However, even for the 
5\% of collisions with highest multiplicity, the near tip-to-tip category remains 
only about 21\% of the events with a low spectator count (the latter being defined 
as the lowest 1\% of all collisions, regardless of multiplicity).  
%%%%%%%%%%%%%%%%%%%%%%%%%%%%%%%%%%%%%%%%

\subsection{With smearing}
In the previous section, we defined ranges of U+U collision configurations 
termed ``near tip-to-tip" and ``near body-to-body".  These ranges were defined 
for illustrative purposes only; in this section, we estimate the characteristics 
of a class of events intended to have the highest experimentally obtainable 
enrichment of the desired tip-to-tip configurations.  We assume that this is 
carried out by selecting events with the highest charged particle multiplicity 
in a broad $\eta$ region centered on midrapidity, and simultaneously cutting 
to select events with a low signal in both Zero Degree Calorimeters.  

The measured $dN_{ch}/d\eta$ in a detector like STAR is linearly correlated 
with the true $dN_{ch}/d\eta$, but is subject to random event-to-event 
fluctuations in the efficiency for detecting tracks as well as random 
variations in the inclusion rate for tracks that do not originate from the 
primary collision vertex.  Similarly, the ZDC signals have a component that is 
proportional to the number of emitted spectator neutrons as well as a component 
from various sources of background.  Furthermore, the RHIC intersection regions  
are arranged such that the ZDCs have acceptance for only free neutrons and an 
insignificantly small fraction of neutron-rich charged spectator fragments.  
Thus while we compute the total number of spectator nucleons in a given event 
using our Monte Carlo Glauber model, in real collisions at STAR, only a subset 
of the spectators are detected by the ZDCs, and so events with a small ZDC 
signal include a high background level of collisions with incomplete overlap.  

\begin{figure}[h]
\begin{center}
  \resizebox{7cm}{10cm}{\includegraphics*{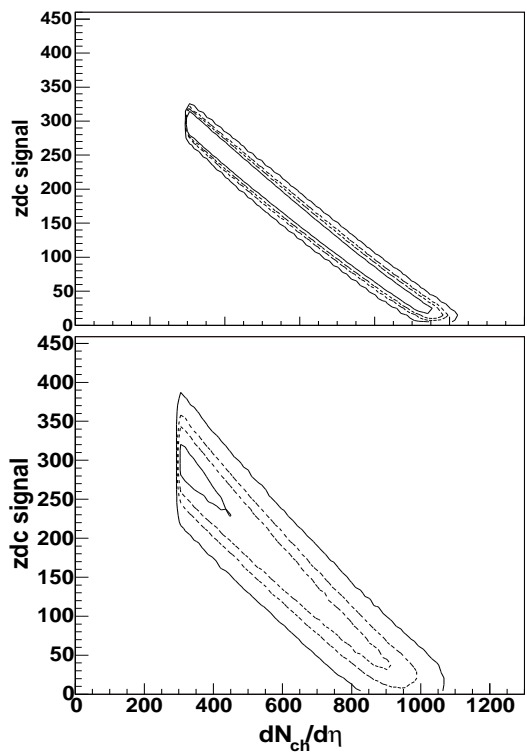}}
\caption{
Contours of event density in the plane of simulated ZDC signal versus 
$dN_{ch}/d\eta$ before (upper panel) and after (lower panel) simulation  
of detector-related smearing and background.  The contour levels are 
the same in both panels, with a ratio of 1:7 between the two outermost 
contours, and a ratio of 1:2 between all other adjacent contours.   
\label{fig:ZDCvsMult}}
\end{center}
\end{figure}

\begin{figure}[h]
\begin{center}
  \resizebox{7cm}{5cm}{\includegraphics*{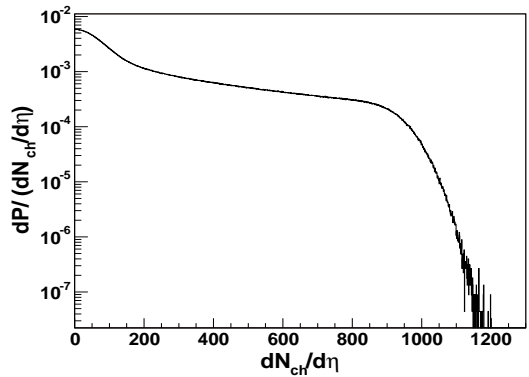}}
\caption{
Charged particle multiplicity distribution in U+U collisions after detector smearing.
\label{fig:mult-smeared}}
\end{center}
\end{figure}

The upper panel of Fig.~\ref{fig:ZDCvsMult} shows contours of event density in 
the plane of spectator count versus $dN_{ch}/d\eta$ from our Monte Carlo Glauber 
model without any detector simulation.  The resulting correlation is much narrower 
than what is observed in experiment \cite{STAR-NIM}, an expected outcome given the 
simplicity of the model and the neglect of the experimental effects discussed above.  
To make a realistic estimate of the best enrichment of tip-to-tip U+U collisions 
that might be achieved, it is necessary to smear out the narrow ridge in the upper 
panel of Fig.~\ref{fig:ZDCvsMult} so that a pattern resembling the experimental 
distribution is obtained.  For this purpose, a Gaussian-distributed random number 
is first added to $dN_{ch}/d\eta$ for each event, such that the steeply dropping 
upper tail of the $dN_{ch}/d\eta$ distribution in Fig.~\ref{fig:mult-no-smear} is 
smeared out to the point where it resembles the experimental distribution from STAR.  
The smeared multiplicity distribution is shown in Fig.~\ref{fig:mult-smeared}.
Note that we are concerned here with the upper end of the multiplicity 
distribution (highest values of $dN_{ch}/d\eta$), and the variation of the smearing 
at lower multiplicities is not of interest.
  
Next, another Gaussian-distributed random number (with non-zero mean) is added to 
the spectator count and the result is used as the simulated ZDC signal.  The 
Gaussian mean and width are adjusted to produce a ridge near the lower right part of 
the lower panel of Fig.~\ref{fig:ZDCvsMult} that resembles the observed data from 
STAR \cite{STAR-NIM}. This procedure has a small effect on the agreement of 
the calculation with the PHOBOS data shown in Fig.~\ref{fig:KNpar}, and while the effect is minimal,
we corrected for the difference.
\begin{figure}[h]
\begin{center}
\resizebox{6.8cm}{5cm}{\includegraphics*{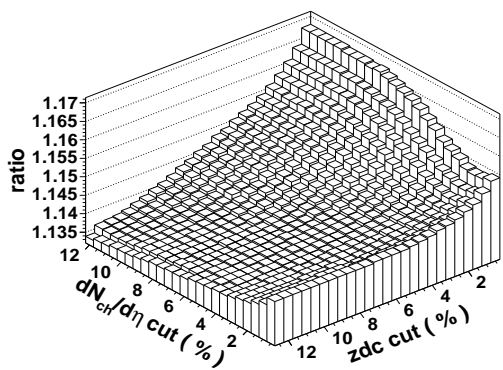}}
\caption{
Ratio $[(1/S)\, dN_{ch}/dy]_{\rm UU} / [(1/S)\, dN_{ch}/dy]_{\rm AuAu}$ as a function 
of different $dN_{ch}/d\eta$ and ZDC cuts, labeled according to the selected 
percentage of the total sample.  
\label{fig:Lego}}
\end{center}
\end{figure}

Various combinations of cuts in $dN_{ch}/d\eta$ and in ZDC signal have been applied 
to the simulated U+U data plotted in the lower panel of Fig.~\ref{fig:ZDCvsMult}
with the objective of maximizing the transverse particle density  
$(1/S)\, dN_{ch}/dy$.  Because the quantity we seek to maximize is trivially 
correlated with one of our cut variables, we assess the maximum 
particle density in U+U collisions relative to Au+Au collisions with the same 
cuts on a percentage basis.  

The vertical axis in Fig.~\ref{fig:Lego} shows the ratio $[(1/S)\, dN_{ch}/dy]_{\rm UU} 
/ [(1/S)\, dN_{ch}/dy]_{\rm AuAu}$ as a function of independent scans across both cut 
variables.  Note the zero-suppressed vertical axis; we conclude that no further 
increase in $[(1/S)\, dN_{ch}/dy]_{\rm UU}$ beyond about 18\% relative to Au+Au 
can be reached in the context of our simulation, even searching beyond the plotted 
combinations of cuts on $dN_{ch}/d\eta$ and ZDC signal.  In the region of the most 
selective combination of centrality cuts, the advantage of U+U is about 16\%.

\begin{figure}[h]
\begin{center}
\resizebox{8.6cm}{3.4cm}{\includegraphics*{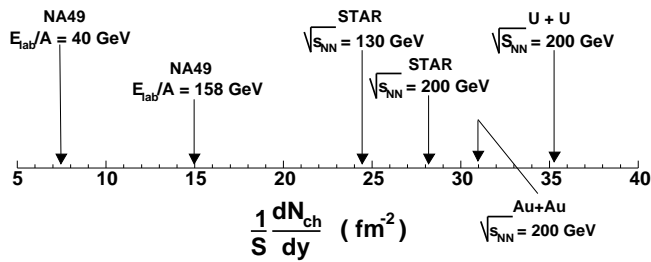}}
\caption{Maximum values of $(1/S)\, dN_{ch}/dy$.  Values from NA49 and STAR correspond 
to central Pb+Pb \cite{Alt:2003ab} and central Au+Au \cite{Adams:2005dq}, respectively.  
The results labeled Au+Au $\sqrt{s_{NN}}$ = 200 GeV and U+U $\sqrt{s_{NN}}$ = 200 GeV 
correspond to the top 5\% in $dN_{ch}/d\eta$ from the present simulation.  
\label{fig:diff_energy}}
\end{center}
\end{figure}
Figure~\ref{fig:diff_energy} summarizes the measured progression in the maximum 
$(1/S)\, dN_{ch}/dy$ for central Pb+Pb or Au+Au collisions, beginning with a low SPS 
beam energy and continuing to the highest RHIC energy.  On the same number line, we 
indicate that when our simulation for Au+Au is constrained to agree well with the 
value from the STAR collaboration, we predict a potential further increase of about 
4 fm$^{-2}$ for U+U with the same centrality cut.
The calculated value of $(1/S)\, dN_{ch}/dy$ is larger than the STAR result for Au+Au
at 200 GeV because of different parameterization of the nuclear radius and small 
uncertainties in the number of wounded nucleons. However, this has a negligible 
effect on the relative increase from Au+Au to U+U.
%\vspace{1cm}

%%%%%%%%%%%%%%%%%%%%%%%%%%%%%%%%%%%%%%%
\section{Conclusion}
\label{concl}

We have undertaken a study of the potential for U+U collisions to realize an 
increased transverse particle density $(1/S) dN_{ch}/dy$ without increasing the beam 
energy. We note that the present investigation addresses only a limited aspect of U+U 
collisions, namely, we have studied the ``worst case scenario" where only the transverse 
particle density $(1/S) dN_{ch}/dy$ is of interest, and where the charged particle 
multiplicity and the Zero-Degree Calorimeter signals are the only means of 
selecting the desired U+U events.  Previous studies have considered the issue 
of the path-dependent energy loss of partons in the unique participant geometry 
of central U+U collisions \cite{Shuryak:1999by,Heinz:2005xd}, and it is feasible 
for U+U collisions to extend our physics reach in a variety of observables.  

The transverse particle density increases by about 35\% in the ideal limit of tip-to-tip 
U+U collisions compared to Au+Au at zero impact parameter.  However, the practical 
limitation of selecting the U+U samples of interest via conventional measures of 
centrality, as used in RHIC experiments, leads to uranium beams offering a smaller, 
yet still-worthwhile advantage.  Specifically, our simulations suggest that the 
maximum achievable values of $(1/S) dN_{ch}/dy$ at RHIC could be increased from 
the present $\sim 31$ fm$^{-2}$ to about 35 fm$^{-2}$ with U+U collisions.  This 
increase should justify the needed investment of effort and resources at RHIC, 
being of the same order as the measured increase for central Au+Au collisions 
when the beam energy was increased from $\sqrt{s_{NN}}$ = 130 GeV to 200 GeV, 
or just under half the increase between the top CERN SPS energy and $\sqrt{s_{NN}}$ 
= 130 GeV. 
%
%%%%%%%%%%%%%%%%%%%%%%%%%%%%%%%%%%%%%%%%
%
\section{Acknowledgments}
We thank Ulrich Heinz, Anthony Kuhlman, and Peter Levai for extensive discussions. One of 
the authors (GF) acknowledges the support of a Szent-Gy\"orgyi Scholarship of the 
Hungarian Department of Education and the hospitality of the E\"otv\"os University, 
where some of this work was carried out.  This work was supported in part by US DOE 
grants DE-FG02-86ER40251 and DE-FG02-89ER40531.  
%
%%%%%%%%%%%%%%%%%%%%%%%%%%%%%%%%%%%%%%%%

\end{document}